\documentclass[sigconf,authorversion,nonacm]{acmart}
\usepackage{algpseudocode}
\usepackage[vlined,linesnumbered,algoruled,boxed,noresetcount]{algorithm2e}
\usepackage[export]{adjustbox}
\usepackage{listings}
\usepackage{enumitem}% http://ctan.org/pkg/enumitem
\usepackage{fancyvrb}
\usepackage{fancyhdr}
\AtBeginDocument{%
  \providecommand\BibTeX{{%
    \normalfont B\kern-0.5em{\scshape i\kern-0.25em b}\kern-0.8em\TeX}}}

\acmSubmissionID{172}

\begin{document}

%\title{StoryVerse: Towards Dynamic Narrative Planning for Games with Large Language Model-based Character Simulation}

%\title{StoryVerse: Towards Authorable LLM-based Character Simulation with Dynamic Narrative Planning}
%\title{StoryVerse: Towards Authorable Plot-driven Character Simulation with Language Models for Dynamic Narrative Planning}

\title{StoryVerse: Towards Co-authoring Dynamic Plot with LLM-based Character Simulation via Narrative Planning}

%Towards Authorable Plot-driven LLM-based Character Simulation with Dynamic Narrative Planning

%Towards Co-authoring Dynamic Plot using LLM-based Character Simulation and Narrative Planning


\author{Yi Wang}
\affiliation{%
  \institution{Autodesk Research}
  \city{San Francisco}
  \country{USA}}
\email{yi.wang@autodesk.com}

\author{Qian Zhou}
\affiliation{%
  \institution{Autodesk Research}
  \city{Toronto}
  \country{Canada}}
\email{qian.zhou@autodesk.com}

\author{David Ledo}
\affiliation{%
  \institution{Autodesk Research}
  \city{Toronto}
  \country{Canada}}
\email{david.ledo@autodesk.com}

\newcommand{\note}[1]{\textcolor{teal}{[Note: #1]}}
\newcommand{\todo}[1]{\textcolor{red}{[todo: #1]}}
\newcommand{\placeholder}[1]{\textcolor{black}{#1}}

\begin{abstract}
Automated plot generation for games enhances the player’s experience by providing rich and immersive narrative experience that adapts to the player’s actions. Traditional approaches adopt a symbolic narrative planning method which limits the scale and complexity of the generated plot by requiring extensive knowledge engineering work. Recent advancements use Large Language Models (LLMs) to drive the behavior of virtual characters, allowing plots to emerge from interactions between characters and their environments. However, the emergent nature of such decentralized plot generation makes it difficult for authors to direct plot progression. We propose a novel plot creation workflow that mediates between a writer’s authorial intent and the emergent behaviors from LLM-driven character simulation, through a novel authorial structure called ``abstract acts''. The writers define high-level plot outlines that are later transformed into concrete character action sequences via an LLM-based narrative planning process, based on the game world state. The process creates ``living stories'' that dynamically adapt to various game world states, resulting in narratives co-created by the author, character simulation, and player. We present \textit{StoryVerse} as a proof-of-concept system to demonstrate this plot creation workflow. We showcase the versatility of our approach with examples in different stories and game environments. 
\end{abstract}





 \begin{teaserfigure}
   \includegraphics[width=\textwidth]{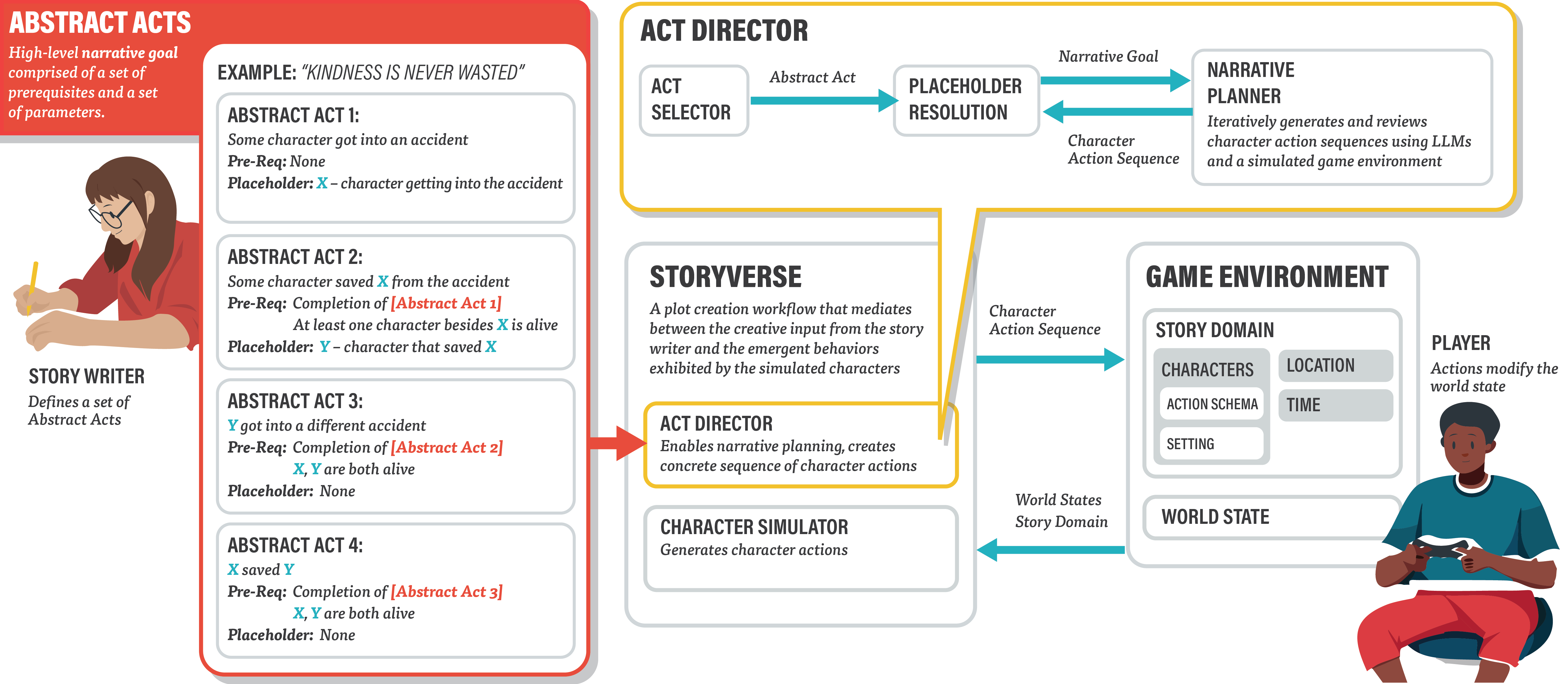}
   \caption{We present a novel plot creation workflow that mediates between a writer's authorial intent and the emergent behaviors from LLM-driven character simulation through an authorial structure called ``abstract acts'' as well as an LLM-based narrative planning process that iteratively
generates and reviews character action sequences. }
   \label{fig:teaser}
 \end{teaserfigure}


\maketitle

%\fancyfoot[C]{\textcolor{red}{AUTODESK CONFIDENTIAL - for internal use only - Submitted to FDG'24 Late-Breaking Short paper track}}

\section{Introduction}

%\note{directable instead of controllable}
%\note{``Plot'' seems a better term than ``story'' in our context}
Automated plot generation for games enhances the player’s experience by providing rich and immersive narrative experience that adapts to the player’s actions. One prominent approach is symbolic narrative planning  \cite{young2013plans, meehan1977tale, lebowitz1985story, riedl2010narrative, ware2011cpocl}, where a narrative goal is set as the plot's end, and character action sequences leading to this goal - known as ``narrative plans'' \cite{lebowitz1985story} - are generated.
Generating such plans requires a hand-crafted knowledge base using formal logical language that defines preconditions and effects of a predefined action set within a story \cite{poulakos2016evaluating}. Given the extensive engineering work required to construct this knowledge base, the generated plots offer limited complexity and scale.

One way to address this challenge is to use Large Language Models (LLMs) to drive character behaviors \cite{park2023generative, wang2023humanoid, chen2023agentverse}. Through LLMs, plots naturally emerge from the interactions between virtual characters and their environments. For example, Generative Agents \cite{park2023generative} simulate believable human behaviors extending LLMs to store a comprehensive log of the agents' experiences, combine their memories to form higher-level thoughts, and retrieve them as needed for behavior planning.
This demonstrates LLM's capabilities of acting as proxies for human behaviors and conversations \cite{inworldorigin}, eliminating the need to craft the knowledge base. 
%, indicating these LLMs already implicitly established a knowledge base sufficient for producing plots. 

Plot generation via fully autonomous characters controlled by LLMs produces {\em emergent narratives} \cite{aylett1999narrative}. However, this type of plot generation \textit{``struggles with `herding cats' when the narrative is required to contain certain content''} \cite{ware2021sabre}.
%While the behaviors of characters can be completely generated by LLMs, 
It is unclear how a story writer might control and direct the plot progression, especially considering how it depends on the collective behavior of multiple characters involved. This is even more challenging for LLM-based emergent narratives due to their black-box nature. 
%Using existing LLM-based character simulation \cite{park2023generative, wang2023humanoid}, writers can enter the initial prompt describing the characters and environments. 
%As the simulation runs, it is unclear how they can direct the collective behaviors of multiple characters - which should be emergent - to adhere to a pre-authored storyline such as ``an evil dragon takes over a village, but then was slayed by a hero''?
%When the plot emerge from the interaction between a set of LLM-driven characters, whose actions are completely taken over by LLMs, how to make sure the plot adheres to a pre-authored structure? 
%While it is possible for writers to directly specify characters' actions by overriding the results from character simulation, this manual way of authoring and revising is tedious, not scalable to multiple characters, not feasible for real-time plot generation, and defeats the purpose of character simulation.
%each with a prolonged list of generated actions. There is no guarantee on where the story is going.
In a more traditional symbolic narrative planning approach, writers are able to set constraints on the generated behaviors, and the planning systems centralize decision making to ensure they meet the author’s constraints \cite{ware2021sabre}. It is unclear what kind of authorial structure could be compatible with emergent narrative from LLM-based character simulation, which is generated in a decentralized manner.

%However, the emergent nature of plots generated with this approach prohibits a human author to direct the progression of the plot in a traditional sense, especially when the plot progression should be the result of multiple character's actions. For example, when the author only has control over the initial description of the characters and the environment, how can they direct the collective behaviors of the characters - which should be emergent - to adhere to a pre-authored storyline such as ``an evil dragon takes over a village, but then was slayed by a hero''?  In the symbolic narrative planning approach for story generation, the author is able to set arbitrary constraints on the generated sequence of character actions to control the result. When the stories emerge from the interaction between a set of LLM-driven virtual characters, character actions are completely taken over by the LLMs. There is no guarantee on where the story is going, and it is unclear what kind of authorial structure could be compatible with this form of emergent storytelling. 
%When the character behaviors are completely driven by the LLMs, with only the initial prompt describing the characters and environments left within control for the authors, 
%It is not even clear what "authoring a story" means in this context. 

In this work, we present \textit{StoryVerse}, a system that uses LLM-based narrative planning and character simulation to generate dynamic stories through a novel authorial structure called ``abstract acts''. \textit{StoryVerse} introduces a plot creation workflow that mediates between a writer's authorial intent and the emergent behaviors from LLM-driven character simulation, given a game environment. Instead of directly specifying characters' actions to interfere with the character simulation, the writer defines high-level plot outlines as abstract acts, which are later ``instantiated'' into concrete acts through an \textit{Act Director} component. When instantiating an abstract act, the \textit{Act Director} takes into account the state of the game world, the previous acts, and character actions resulting from an ongoing LLM-driven \textit{Character Simulator} component. The resulting concrete act contains a sequence of character actions, which can be executed in the game environment to update the world state, thus informing subsequent character simulation and the following acts. This allows writers to create ``living stories'' that dynamically adapt to various game world states, influenced by the writer's authorial intents, character simulation results, and players' actions.
To demonstrate the feasibility of the proposed workflow, we created a proof-of-concept prototype integrating an LLM-based character simulator similar to \textit{AgentVerse} \cite{chen2023agentverse}, and a novel LLM-based narrative planner, tightly coupled by the game environment and the state of the game world. We showcase \textit{StoryVerse} with examples in different stories. Specifically, our  contributions are:

\begin{itemize}[nosep, topsep = 2pt]
    \item A plot creation workflow that mediates between the creative input from the story writer and the emergent behaviors exhibited by simulated characters. 
    \item An authorial structure called ``abstract acts'' that enables co-creating dynamic plot between a writer and an LLM-based character simulation.
    \item An LLM-based narrative planning process that iteratively generates and reviews character action sequences using LLM and a simulated game environment.   
    \item \textit{StoryVerse}: a proof-of-concept prototype that demonstrates the feasibility and applications of the proposed workflow.
\end{itemize}
\vspace{-4mm}

\section{Method: Dynamic Plot Creation using LLM-based Narrative Planning and Character Simulations}

% overview of the workflow: what is the goal, major components of Figure 1, major techniques used in the system
Our work is built on top of the character simulation in game environments where the characters' behaviors are driven by LLMs \cite{park2023generative}. The goal is to provide a novel plot creation workflow that allows a story writer to direct the progression of the plot while leaving room for the details to emerge from the character simulation as well as the interaction between the player and characters. 
In this section, we describe our workflow centered around the concept of ``abstract act'' and LLM-based narrative planning.

\begin{figure*}
    \centering
    \includegraphics[width=\linewidth]{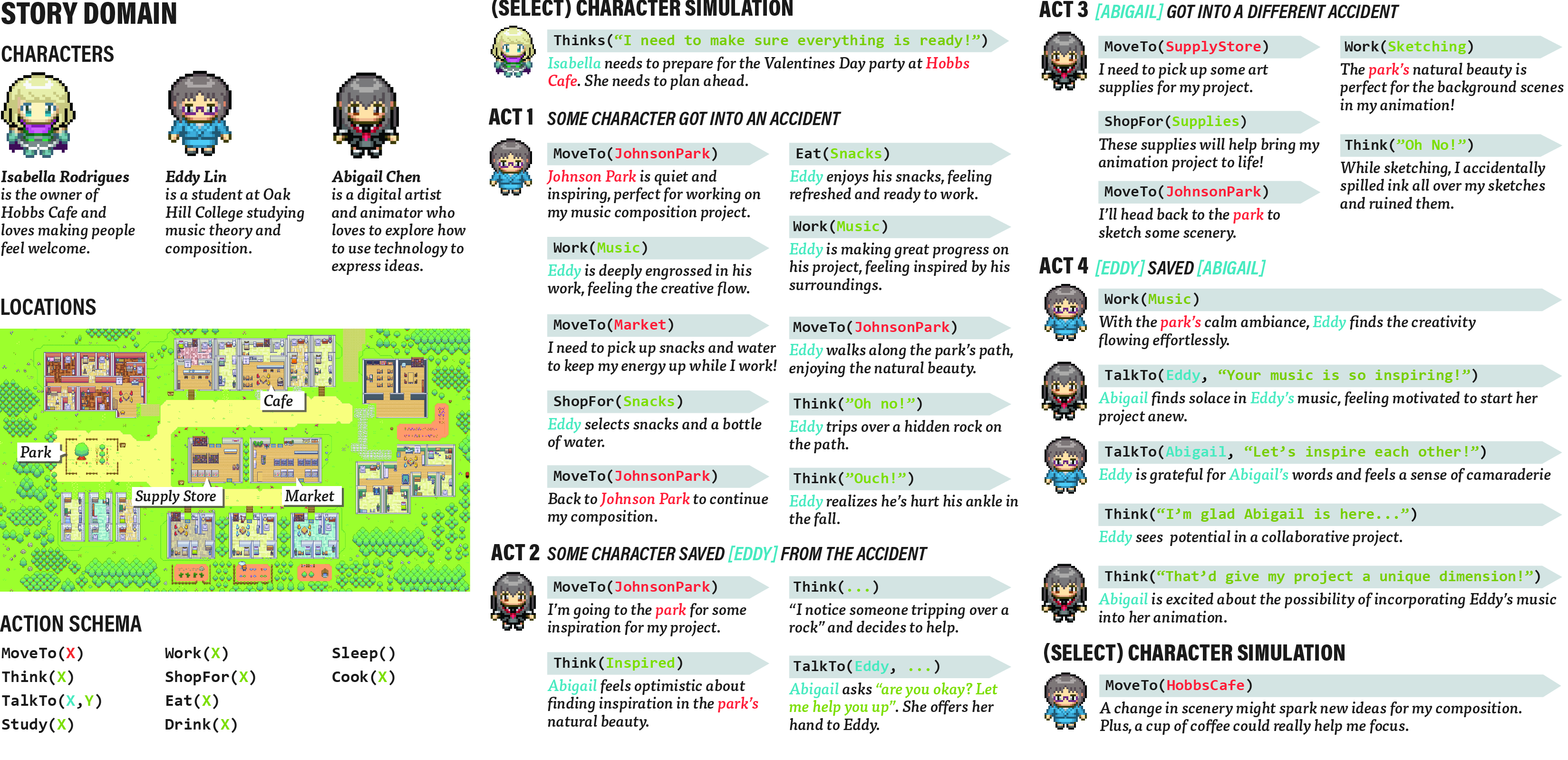}
    \includegraphics[width=\linewidth]{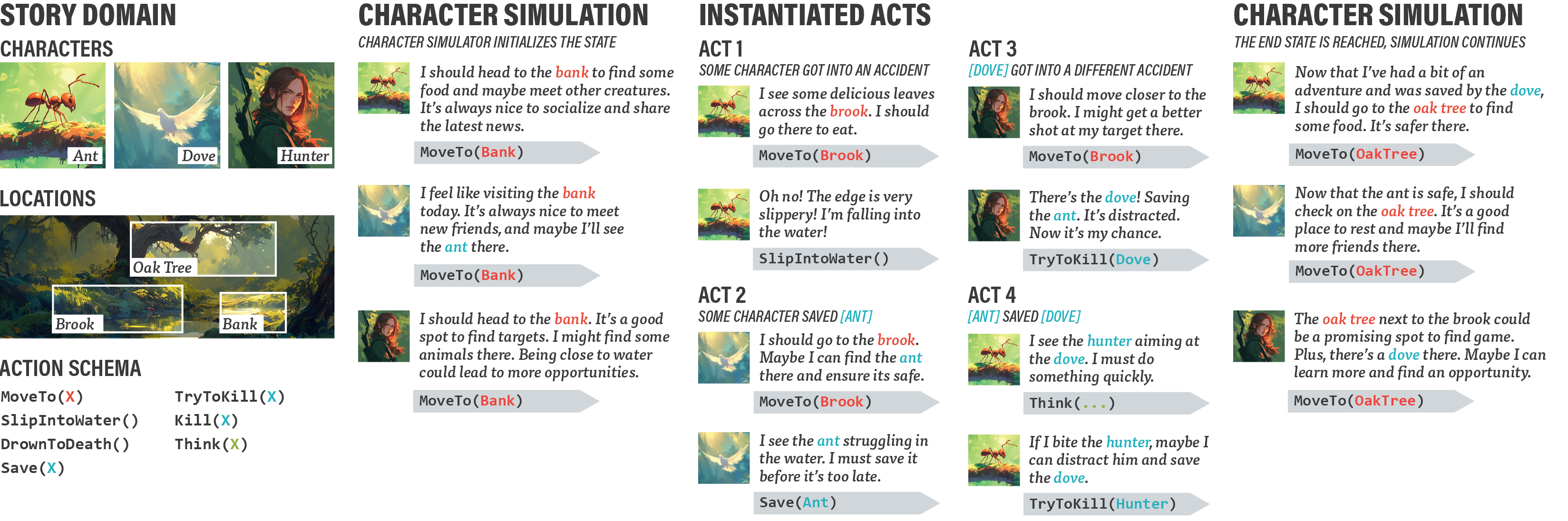}
    \caption{Two example story domains - The Ville (top) and Ant \& Dove (bottom) - together with instantiated versions of the four abstract acts from Figure \ref{fig:teaser}. Note that the text for narrations, dialogs, and monologues is all generated by LLMs.}
    \label{fig:example}
\end{figure*}

\subsection{Plot Creation Workflow}

%StoryVerse presents a new approach to generating stories that combines the input from writers with the simulated behaviors of characters in a game environment using LLM-driven agents. 
Our goal is to allow writers to author abstract yet high-level plot outlines that can be automatically ``instantiated'' into specific character actions at play-time, with the instantiation taking into account the state of the game world.
The system integrates three major components (Figure~\ref{fig:teaser}): an \emph{Act Director}, a \emph{Character Simulator}, and a \emph{Game Environment}. The \emph{Act Director} processes the writer's input and uses an LLM-based narrative planner to generate character action sequences. The LLM-based \emph{Character Simulator} produces character actions that drive the characters by default when no eligible input is found from the writer. The \emph{Game Environment} contains a set of characters, locations, and action schemas that specify character actions that are executable through the game system. Consistent with symbolic narrative planning \cite{poulakos2016evaluating}, we refer to these elements as the {\em Story Domain}. Figure \ref{fig:example} (left) shows two examples of story domains. The game environment also maintains the {\em World State}, which consists of a collection of variables that hold relevant values for the game mechanics, such as the characters': location, attributes (e.g., strength and health points), relationship scores, as well as their memories from the simulation. 
To enable writers to co-develop character actions with the LLM-driven character simulation, we introduce abstract acts, a novel authorial structure compatible with character simulation. The writer can define a high-level plot outline through a set of abstract acts.

At each timestep during play-time, the system evaluates whether there are any pending abstract acts that have met their prerequisites. If there are, the \emph{Act Director} will process these acts and generate character actions. Otherwise, the \emph{Character Simulator} will execute to generate character actions. After the character actions are generated, either from \emph{Act Director} or \emph{Character Simulator}, the \emph{Game Environment} executes the actions and updates the world states. The \emph{Game Environment} ensures that the world states reflect the changes resulting from the generated character actions. Once the game environment executes the actions, the game player can interfere with the game by changing the world states, such as by saving or killing a character. Finally, the game environment sends the updated world states back to \emph{Act Director} for the next timestep. 

This process creates a living story with its progression influenced by the story writer, the character simulator, and the actions of the players within the game environment. While the writer defines abstract acts, which describe high-level dramatic conflicts and turning points, this authorial structure leaves space for the details to emerge from both the character simulation and how the player interacts with the characters. As a result, the stories have the flexibility to evolve and respond to these dynamic factors.

%\todo{remove placeholder from this section}

\subsection{Abstract Acts}

%In this section, we describe the design of  \placeholder{Abstract Acts}. 
In traditional screenplay writing, writers use the concept of an \textit{``act``} to structure a sequence of movements that turns on a major reversal in the value-charged condition of a character’s life \cite{mckee1997story}. Inspired by this notion, we use the term abstract act to structure a sequence of character actions leading toward a (possibly abstract) narrative goal. The term \textit{``abstract''} highlights the separation of the story from the specific details of the underlying world states. This allows for the final story to be a collaborative creation involving the author, the player, and the emergent behaviors from the LLM character simulation. By abstracting the story from the specifics of the world states, it becomes more adaptable and open to contributions from multiple sources, enabling a dynamic storytelling experience.
%The narrative goals are intended to be ``instantiated into'' a concrete sequence of character actions later, with an AI-based narrative planning process. The process is triggered by a set of condition specified by the author. Each abstract act can establish named plot elements that can be referred to in subsequent abstract acts.

An {\em abstract act}
%is a high-level narrative goal with prerequisites and placeholders for certain story elements such as a character, a location, etc., 
is defined by the following elements:
\begin{itemize}
    \item A high-level narrative goal representing a dramatic conflict or a turning point in the story (e.g., \textit{``a character gets into a life-threatening accident''}). 
    \item A set of prerequisites connected by logical conjunction and/or disjunction. Each prerequisite can be:
    \begin{enumerate}
        \item A statement about the current world state that can be evaluated (as true or false) by the LLM or game environment (e.g., `\textit{`John is loved by all characters''} or \textit{``there are at least 5 characters alive''}),
        \item A statement about the player's action, (e.g., \textit{``The player executes \texttt{eat(X)} action with \texttt{X} being a poisonous food item''}), 
        \item The fulfilment or failure of another abstract act, used to specify dependencies between abstract acts.
    \end{enumerate}
\item A set of placeholders -- a construct representing an answer to be fulfilled at a future time (similar to "Promises" in computer programming). Placeholders refer to specific content in the ``instantiation'' of current act, to carry over to subsequent acts. Each placeholder is a pair of code-name and natural language description (e.g., \textit{``X - the character who got into the life-threatening accident''}). 
%The natural language description can possibly contain the code-name of other placeholders from a different abstract act. \
\end{itemize}

The narrative goal and prerequisites may include placeholders for story specifics that will be determined by the instantiation of a previous act. For example, \textit{``a character saves X's life''}, with \textit{``X''} being a placeholder for the character who \textit{``got into a life-threatening accident''} in a previous act (which is unknown at the time of authoring). \footnote{Note that the occurrence of a placeholder in a prerequisite implies the completion of an act with the placeholder as an output as another prerequisite.}
Incorporating placeholders into the narrative goal and prerequisites allows the story to adapt and respond to the choices and actions made by players as well as the interventions from character simulation. In addition, it enforces coherence of plot elements across abstract acts and allows the author to define continuations between acts without knowing the specifics of each act in advance.

When multiple abstract acts are added by the writer, they are not executed sequentially. Each act has its own prerequisites as defined by the writer. This approach allows the writer to create branching stories, where the path to take is determined by both the player's actions, as well as the character simulation. 
Additionally, abstract acts can be grouped by the writer, indicating that they form a complete storyline. This means that the entire story reaches completion only when all the abstract acts within the group have been executed. Grouping abstract acts allows for a structured narrative authoring experience.

\begin{example}
\label{eg:kindness_is_never_wasted}
We use a story about \textit{``Kindness is never wasted''} to showcase the usage of abstract acts extracted from the Ant \& Dove story from Aesop Fable following prior work on narrative planning \cite{GDC-work, graesser1981incorporating}. As shown in Figure~\ref{fig:teaser}, the writer creates the four abstract acts with high-level narrative goals such as life-threatening and life-saving events to depict the turning points as well as the relationship between characters. Note that the execution order does not depend on the creation order. The placeholders for characters indicate the chronological relationship between acts and enforce consistency among them. 
\end{example}

%\todo{Is the player character taken into account in this narrative planning framework?}
%\todo{"action" equivalent to "beats"}
%\note{Need to distinguish between the ``story'' co-created by the author and the character simulation, and the ``absract story'' the author is defining}

\subsection{Act Director} \label{sec:narrative_planning}

Using the writer-specified abstract acts, the Act Director enables a narrative planning approach using LLMs to generate a concrete sequence of character actions that can achieve the narrative goals specified in the abstract acts. The sequence of character actions is called a {\em (narrative) plan}.

At each timestep of play-time, the Act Selector evaluates whether there are any pending abstract acts that have met their prerequisites and selects a qualified act to execute. 
%\todo{please review and choose a better way to describe}
Once selected, the abstract act undergoes a pre-processing step, where the placeholders for specific plot content contained in the narrative goal are replaced with the actual referred content established and saved in the execution of the previous abstract acts (if any). 
%the Placeholder Resolution component replaces each Placeholder for specific plot content defined in the current abstract act with the actual referred content, according to a mapping between the Placeholders and referred content. 
For example, a narrative goal  \textit{``Y saved X from the accident''} might become \textit{``The dove saved the ant from the accident''}. 
%Note that ``another character'' is not replaced with a concrete character since it is not resolved or saved in the previous acts. 

\begin{figure}
  \centering
  \includegraphics[width=\linewidth]{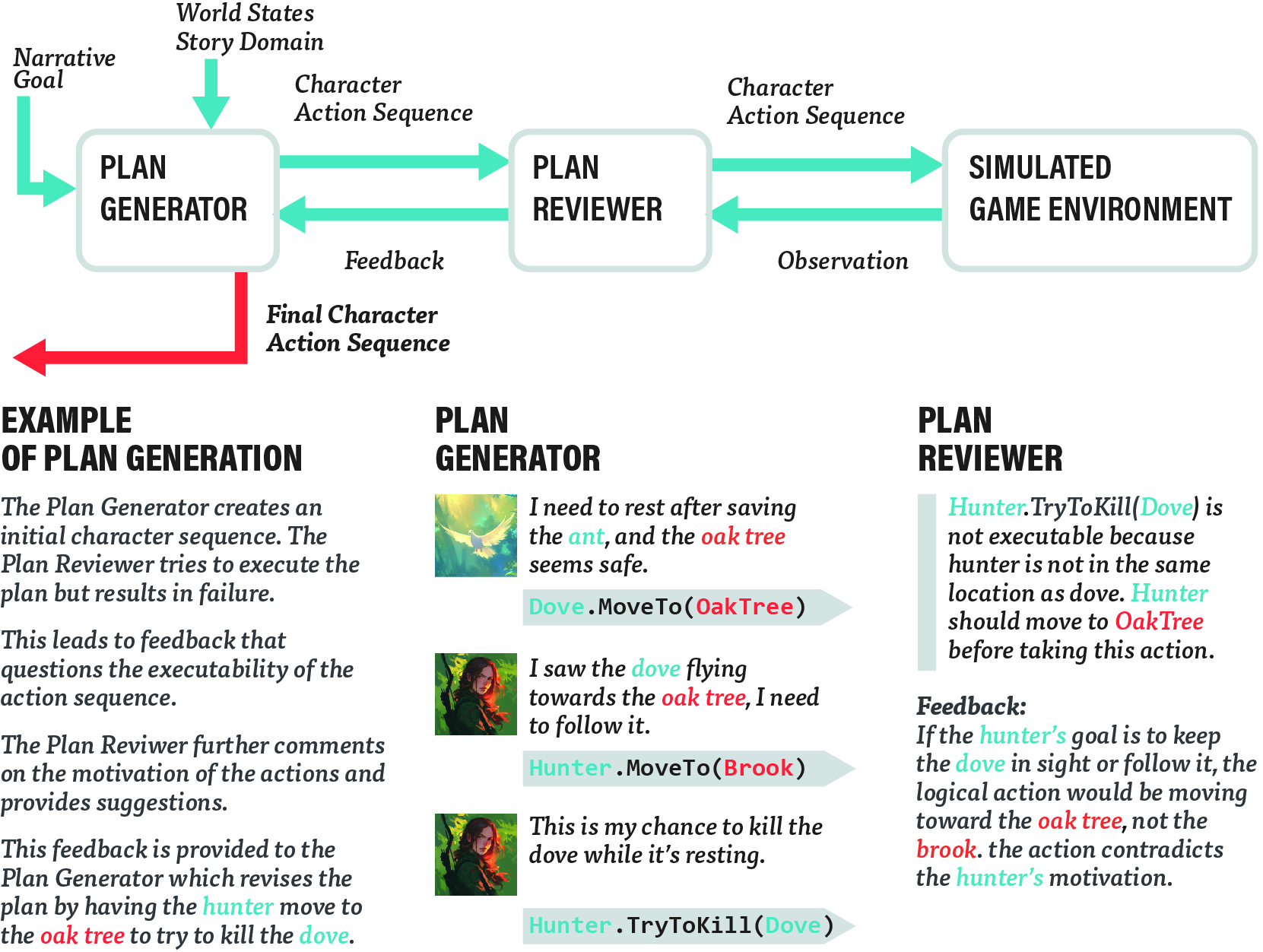}
  \caption{LLM-based narrative planning that iteratively generates and reviews the plan.}
  \label{fig:narrative_planning}
\end{figure}

Following \cite{wang2023describe}, we design an LLM-based iterative narrative planning process involving plan generation and plan revision for better quality of the narrative plans (Figure~\ref{fig:narrative_planning}). 
%Algorithm \ref{alg:narrative_planning} describes the process, where $plan_{action}$ denotes the plan obtained by truncating $plan$ so that $action$ is the last action. 
The plan generator takes a narrative goal, the world state, and the story domain as the input and is prompted to generate a plan.  Once generated, the plan reviewer provides feedback regarding the quality and feasibility of the action sequence to improve it. The feedback has three parts:

\begin{itemize}
    \item {\bf Overall Coherency Evaluation} Feedback is obtained by prompting an LLM to comment on the overall coherency of the generated plot and make suggestions for improvement.
    \item {\bf Game Environment Evaluation}\hspace{2mm} Similar to \cite{wang2023describe}, feedback comes from executing the generated action sequence in a simulated game environment, and reporting observations on the success/failure of the execution. 
    \item {\bf Character Simulation Evaluation}\hspace{2mm} For every action in the sequence, we prompt an LLM to play the role of the subject of the action. Given the current world state including the character's memory, we ask the LLM if the motivation for the character to perform the action has been established. We include the explanation to this question in the feedback, if the motivation has not been established.
    \footnote{In addition, we also ask an LLM if the narrative goal is already met at the current action. If the answer is yes, the generated plan will be truncated to have the current action as the last action. }
\end{itemize}

The plan generator is then prompted to revise the previous action sequence according to the feedback received. The process is repeated for a user-specified number of times, or until the generated plan is executable.

%This process enables an iterative plan refinement process using LLM and the game environment to evaluate the quality of the generated action sequence. Compare to symbolic planning approach,  

Once the final sequence of character actions is generated, it will be executed by the game environment to update the world state. Meanwhile, it will also be sent to the Placeholder Resolution component (Figure~\ref{fig:teaser}), which prompts an LLM to identify  specific content referred to by each placeholder from the current act, and store the placeholder-content mapping for future act execution. For example, if placeholder $X$ is defined to refer to ``the character who got into an accident'', then Placeholder Resolution will ask an LLM, ``what is the character who got into an accident'', and then store the answer with placeholder $X$.

\begin{example}
The generated narrative plans are in the form of a list of tuples containing an action following the action schema in the story domain, the subject of the action, and the thought behind the action, for example:

\begin{small}
\begin{verbatim}
{ 
    "subject": "dove",
    "action": "moveTo(oak_tree)",
    "thought": "I sense danger!
                I must fly away quickly to avoid the hunter."
}
\end{verbatim}
\end{small}
Figure \ref{fig:narrative_planning} shows an iteration of plan revision for Act 3 as described in Figure~\ref{fig:teaser}. The plan was generated using the Ant \& Dove story domain from Figure \ref{fig:example}.

%\includegraphics[width=\linewidth,left]{figures/PlanGenerationFinal.jpg}

% The plan reviewer attempts to execute the plan in the game environment resulted in failure. This leads to feedback that questions the executability
% of the action sequence: 
% \begin{verbatim}
% hunter.tryToKill(dove) is not executable because hunter
% is not at the same location as dove. hunter should move to
% oak_tree before taking this action.
% \end{verbatim}

% The plan reviewer further comments on the motivation of the actions and provides suggestions as: 
% \begin{lstlisting}
% If the hunter's goal is to keep the dove in sight or follow it, the logical action would be moving towards the oak_tree, not the brook. Therefore, the action contradicts the hunter's stated motivation.
% \end{lstlisting}

% This feedback is provided to plan generator, which revised the plan by having the hunter moving to the oak tree and try to kill the dove there.
\end{example}

\subsection{Implementation}
For demonstration purposes, we implemented a basic LLM-based character simulation component similar to a minimal version of \textit{AgentVerse} \cite{chen2023agentverse}. Each character is associated with a pre-authored textual description, and a structured memory storage updated upon execution of actions. At each timestep, each character takes an action generated by an LLM. The component can be replaced with a more sophisticated implementation such as \cite{chen2023agentverse, park2023generative, wang2023humanoid}. 
We also implemented a proxy for game environment which simply maintains a list of variables defining a world state, and updates the variables upon execution of character actions. We use prompt chaining techniques to achieve complex tasks by dynamically assembling multiple prompts. Hand-crafted few-shot examples are used in the prompts to increase the quality of outputs.
We used gpt-4-0125-preview as the LLM for all the experiments reported in the next section. The maximum number of narrative plan revisions ($maxStep$) was  $2$.

%\note{Blue boxes are components involving LLM querying}

%\todo{We will release the code on github}

%\section{Evaluation}
%\note{Probably gonna remove this section due to short paper}

\section{Examples}
We describe two examples to demonstrate plot creation workflow using two story domains: {\it The Ant \& Dove} and {\it The Ville}, which have been previously used in narrative planning \cite{GDC-work, graesser1981incorporating} and generative agents \cite{park2023generative} research. Figure~\ref{fig:example} lists the characters, locations, and action schema for these story domains. Both examples are driven by a set of {abstract acts} based on the theme of "Kindness is never wasted" (Figure~\ref{fig:teaser}). Before any acts are executed, an initial round of character simulation sets the stage for the story by simulating the daily lives of the characters (Figure~\ref{fig:example} character simulation). The Act Director then takes the four abstract acts as input and generates the action sequence for each act in both story domains. 

\begin{figure}
  \centering
  \includegraphics[width=\linewidth]{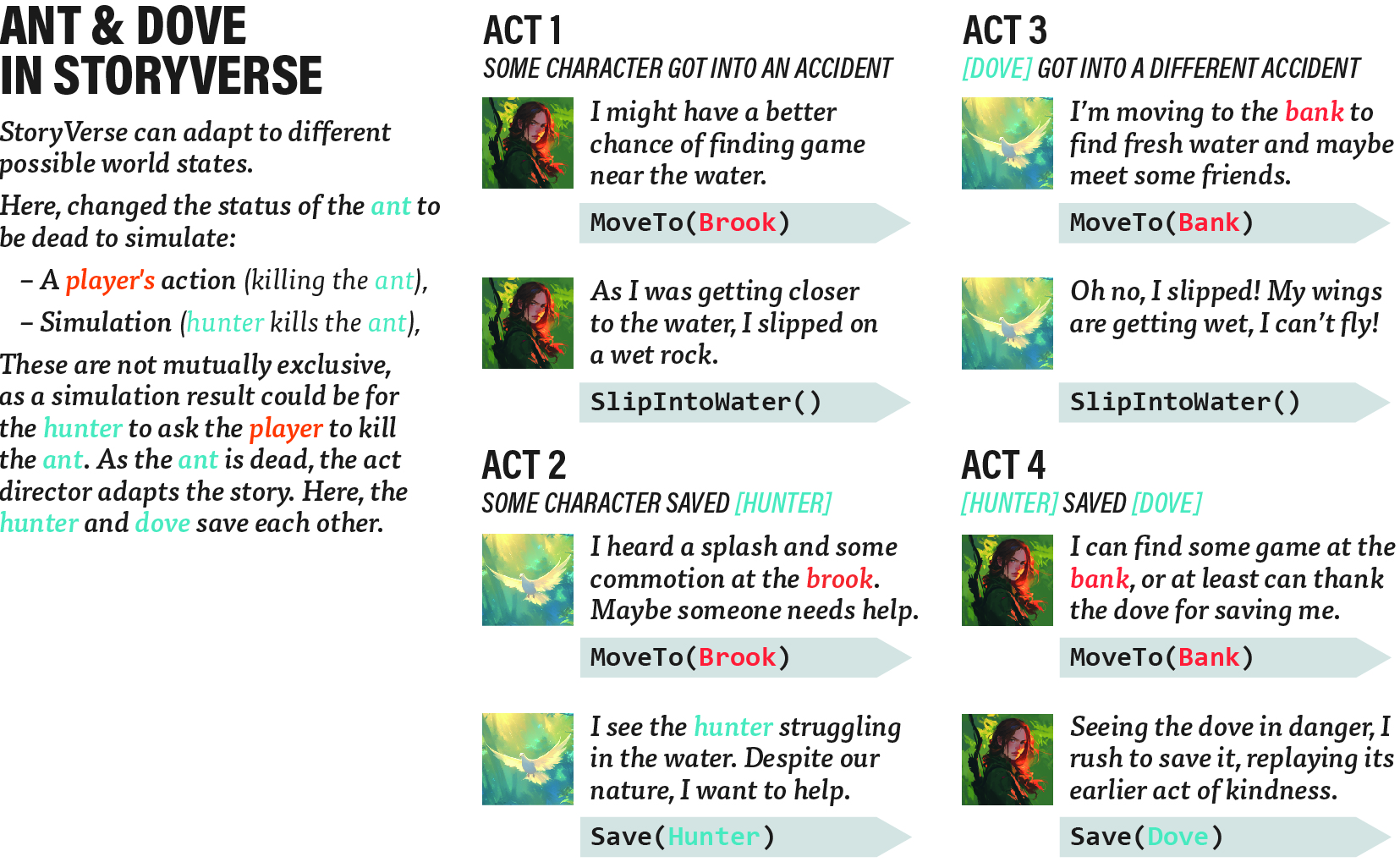}
  \caption{A variation of Ant \& Dove story on ``A kindness is never wasted'' instantiated with the ant dead.}
  \label{fig:a_kindness_is_never_wasted_ant_and_dove_no_ant}
\end{figure}

\begin{example}
In {\it Ant \& Dove} (Figure~\ref{fig:example}, bottom), the ant (character X) slips into the water in Act 1 while searching for food, which continues its behavior from the character simulation. The dove (character Y) rescues the ant in Act 2. Although the hunter character is not explicitly mentioned in the abstract acts, the narrative planner includes this character as part of the second accident that endangers the dove in Act 3. Finally, the ant saves the dove from the hunter in Act 4, completing the recreation of the original {\it Ant \& Dove} fable. After the four acts conclude, the characters return to their daily routines through the character simulation. The actions performed by the characters in the post-character simulation reflect the events that took place during the four acts.

Additionally, to demonstrate how the story adapts to different world states, we manually change the status of the ant to be dead. 
%This could be simulating a result of the player's action (e.g., the player killed the ant), or character simulation (e.g., the hunter killed the ant), or both (The hunter asked the player to kill the ant). 
As the ant is dead, the Act Director adapts the story to the available characters shown in Figure~\ref{fig:a_kindness_is_never_wasted_ant_and_dove_no_ant}. 
\end{example}

\begin{example}
We demonstrate the transferability of abstract acts across story domains using the same set of abstract acts in {\it The Ville}. The story (Figure~\ref{fig:example}, top) includes three characters: Isabella (cafe owner), Eddy (music student), and Abigail (digital artist). Eddy plays character X, who injures his ankle in an accident (Act 1). Abigail plays character Y, who saves Eddy and offers help (Act 2). Abigail later experiences demotivation after accidentally ruining her creations (Act 3), but Eddy's music inspires her to regain motivation and work on her sketches again (Act 4). Note that Isabella, although introduced as a character, is not selected to participate in the acts due to her busy schedule with the Valentine's Day party at the cafe, which prevents her from leaving and interfering with the story unfolding in an outdoor park.
\end{example}

In our examples, the four acts happen to be executed sequentially with no character simulation in between. In other cases, acts can be executed interlacing with character simulation, likely due to the prerequisites not immediately satisfied after the previous act, such as waiting for a few timesteps of character simulation. 

%\todo{Restart mechanism}

%\todo{Show two plots generated differently due to simulation (probably no space for this)}

\section{Discussion}
We present our \textit{StoryVerse} system as a proof-of-concept to a plot creation workflow that mediates between a writers’ authored content and the emergent behaviors from character simulation. For practical use of this approach in games, further technical aspects need to be considered. Like other work utilizing LLMs for story generation, the quality of the generated plots is limited by the LLMs' known challenge of preserving long-term dependency and coherence \cite{mirowski2023co}. It is worth investigating how to incorporate existing solutions such as increasing the size of LLM's context window \cite{kaddour2023challenges}, utilizing Retrieval-Augmented Generation \cite{lewis2020retrieval} and adopting a hierarchical generation approach \cite{mirowski2023co} to mitigate this issue. The frequent call to LLMs also raises concerns about delays in the act execution process during gameplay.

%\textit{StoryVerse} is an experimental system for building interactive narratives. 
Riedl and Bulitko \cite{riedl2013interactive} have proposed a taxonomy of  interactive narrative systems in terms of three dimensions - authorial intent, character autonomy, and player modeling. Within this design space, \textit{StoryVerse} situates at hybrid authorial intent with a relatively strong level of virtual character autonomy, which has been identified as a gap in \cite{riedl2013interactive}. The Act Director serves as an autonomous surrogate for the writer, which has a similar role as the experience manager in \cite{riedl2008dynamic}. However, an LLM-based approach enables a more flexible and generalizable representation of author's constraints on the plot search space (such as the ``island'' concept from \cite{riedl2008dynamic}), and reduces the required knowledge engineering expertise for specifying authorial intents compared to symbolic planning based approaches. 
The work can be further extended in the player modeling dimension. Our proof-of-concept system simulates the player intervention via updating the world states caused by player actions. Potential Improvement in player modeling could result from incorporating a persona derived from the player's preference and the protagonist's character role. For example, when instantiating abstract acts into concrete plots, involving characters preferred by the player/protagonist would likely enhance the immersive experience. 

\section{Conclusion}
In this work, we proposed a novel plot creation workflow intended to balance authorial intent and emergent behaviors for a game world with LLM-driven virtual characters. The workflow allows stories to be co-created by the author, the character simulation and the player's actions asynchronously. We present our prototype system \textit{StoryVerse} as a proof of concept to demonstrate the feasibility of proposed workflow with examples showcasing the system's potential.
Our next steps include systematic evaluations in terms of standard quantitative metrics for story generation, as well as user studies, for us to understand the performance of the technical approach, and the feasibility and potential of the workflow in a real-world setting. We see \textit{StoryVerse} as an early step towards novel interactive narrative design workflows that leverage LLMs to enable immersive narrative experiences tightly coupled with story and game mechanics.

% \begin{acks}
% \end{acks}

\bibliographystyle{ACM-Reference-Format}
\bibliography{base}


%\newpage
%\appendix
%\include{appendix}

\end{document}